\documentclass[usenatbib,useAMS,usegraphicx]{mn2e}

 \usepackage{times}
 \usepackage{amsmath}
 \usepackage[usenames,dvipsnames]{color}

  \definecolor{brown}{RGB}{165,42,42}

  \newcommand{\cmd} {~cm$^{-2}$}
  \newcommand{\cmt} {~cm$^{-3}$}
  \newcommand{\s} {~s$^{-1}$}
  \newcommand{\e}[2] {#1\times10^{#2}} 

  \newcommand{\hd} {H$_2$}
  \newcommand{\hdo} {H$_2$O}
  \newcommand{\hdco} {H$_2$CO}
  \newcommand{\ndh} {N$_2$H$^+$}
  \newcommand{\nht} {NH$_3$}

\begin{document}

\title[Effects of H$_{\rm 2}$ coating of grains on depletion]{Effects of
  H$_{\rm 2}$ coating of grains on depletion of molecular species}

\author[O.~Morata and T.~I.~Hasegawa]{
  Oscar Morata\thanks{e-mail:omorata@asiaa.sinica.edu.tw} and Tatsuhiko I. 
  Hasegawa  \\ 
  Institute of Astronomy and Astrophysics, Academia Sinica, P.O.\ Box 23-141,
  Taipei 106, Taiwan
}

 \maketitle

\begin{abstract}
  Physical conditions in dense and cold regions of interstellar clouds favour
  the formation of ice mantles on the surfaces of interstellar grains.
  It is predicted that most of the gaseous species heavier than \hd\ or He
  will adsorb onto the grains and will disappear from the gas-phase, changing
  its chemistry, within $\sim10^9/n_{\rm H}$ years.
  Nonetheless, many molecules in molecular clouds are not completely depleted
  in timescales of 10$^5$ yr.
  Several speculative mechanisms have been proposed to explain why molecules
  stay in the gas phase, but up to now none are fully convincing.
  At the same time, these mechanisms are not mutually exclusive and we can
  still explore the effects of other possible processes.
  We speculate on the consequences of H$_2$ coating of grains on the
  evaporation rates of adsorbed species.
  More experiments and simulations are needed to calculate the evaporation
  rate $E_{\rm evap}$(X-\hd).
\end{abstract}


\section{Introduction}

  Icy mantles on dust grains have been long detected in cold and den\-se
  regions of the interstellar medium \citep{Williams92}.
  These mantles are the result of the accumulation of heavy molecules on grain
  surfaces due to the low temperature and high density conditions of the
  environment.
  The collisions between the dust grains and the neutral component of the gas
  lead to the retention of the gas phase particles on the grain surface,
  because their thermal energy is lower than the typical adsorption energies
  and because the excess kinetic energy can be transferred rapidly to the
  surface \citep{HerbstMillar91}.
  Calculations have shown \citep{LeitchDevlin85} that $\sim30$--100\% of the
  collisions result in retention, depending on the species and the nature of
  the surface.
  Moreover, the widespread detection of molecular mantles in dense regions of
  molecular clouds indicates a high efficiency of the sticking of heavy atoms
  and molecules on the surfaces of cold dust grains.

  The main observed component of the mantles is water ice
  \citep{TielensHagen82,Whittet98}, but it is thought that most of the
  gaseous species heavier than \hd\ or He will adsorb onto the grains and will
  disappear from the gas-phase, changing its chemistry, in a relatively short
  ``freeze-out'' time.
  The freeze-out rate is $\sim10^9/n_{\rm H}$ years \citep{LeitchDevlin85},
  where $n_{\rm H} = n({\rm H}) + 2n({\rm H_2})$ is the total hydrogen nucleon
  number density.
  For densities as low as $\sim10^4$\cmt, \textit{the freeze-out timescale is
    much less than the expected lifetime of a typical molecular cloud
    ($\sim10^7$ yr)}, so it would be expected that, contrary to the
  observational result, the majority of observations would not show evidence
  for heavy gas-phase species.
  \citet{Iglesias77} showed that for non-zero sticking efficiency, when no
  desorption processes act to remove mantle material, the accretion of
  gas-phase species on to the dust dominates the chemical evolution and the
  total freeze-out of gas is inevitable.

  Thus, some mechanism is preventing the growth of mantles in grains and
  maintains an equilibrium between molecules in the gas and solid phases by
  producing the desorption of molecular species from the surface of dust
  grains.
  The finding of a convincing desorption mechanism that explains the observed
  molecular gas-phase abundances has proved to be rather elusive in the last
  20 years.
  It is not yet precisely known how this desorption takes place and several
  different continuous desorption processes have been proposed, either working
  simultaneously or separately \citep{Roberts07}:

  \begin{itemize}
    \item[-] Classical thermal evaporation \citep{Leger85}, which is
      negligible at the temperatures of dense clouds, $\sim 10$~K

    \item[-] desorption by direct impact of cosmic rays (or X-rays) onto
      grains, which causes local heating that lead to the evaporation of
      weakly bound molecules, such as CO \citep{Willacy93,Leger85,Hasegawa93}

    \item[-] cosmic ray-induced photo-desorption \citep{Hartquist90,Willacy94}

    \item[-] exothermic mantle reactions, such as the formation of H$_2$,
      which can result in a local ``hotspot'' heating of the grain mantle
      \citep{WillacyWillDul94,DuleyWilliams93}

    \item[-] grain-grain collisions or chemical explosions \citep{Langer00};
      grain-grain collision or sputtering in interstellar shocks
      \citep{Tielens94}

    \item[-] if the cloud is composed of clumpy molecular gas layers,
      interstellar UV radiation can penetrate two or three times deeper than
      in the case of a homogeneous cloud and can produce orders of magnitude
      more photo-ionisation and photo-desorption even at modest extinctions,
      $A_V\sim2$ \citep{Boisse90,Bethell07}.
  \end{itemize} 

  The feedback from the star formation processes, which eventually disrupt the
  dense cores from which stars form, will also limit the growth of ice mantles
  or force the return of the adsorbed species to the gas phase, in a sporadic
  or intermittent manner, through a variety of mechanisms:
  by sublimation of icy mantles from warm grains near bright stars in a more
  or less gradual way \citep{VitiWilliams99,Schoier02,Rodgers03,Garrod08}, by
  destruction or sputtering of grain cores and mantles by passing shocks
  \citep{Charnley88,Flower94,Bergin98,vanDishoeckBlake98}, or by ice
  sublimation due to UV radiation originating in shocks
  \citep{Viti03,Christie11}.

  All the mechanisms above are not mutually exclusive, but no proposed
  non-thermal mechanism convincingly explains the pre\-sence of molecules such
  as CO in dark clouds or it is widely accepted as a reliable desorption
  mechanism \citep[e.g.,][]{Frerking82,Duvert86,DickmanHerbst90}.
  In this paper, we propose to explore another possible mechanism:
  the con\-se\-quences of the variation of the adsorption energies due to the
  different properties of the substrate and, in particular, the effect that
  the coating of dust grains by molecular hydrogen has on the evapo\-ration
  rates of adsorbed molecules.

\section{Effects of H$_2$ coating of grains on adsorption energies}

  Once a molecular species sticks to a grain, the time it stays on the surface
  before it evaporates depends on the surface temperature and the binding
  (adsorbing) energy of the adsorbate species on a given substrate.
  The desorption rate, in\s, is described by \citet{Hasegawa92}

  \begin{equation}
   R_{\rm evap} = \nu_0 ~{\rm exp} (-E_D / T )
  \end{equation}
  where $T$ is the surface temperature of the grain, $E_D$ the adsorption
  energy in K, and $\nu_0$ is the characteristic vibrating frequency for the
  adsorbed species, $\nu_0=(2n_sE_{\rm D} / \pi^2m)^{1/2}$, where $n_s$ is the
  surface density of sites ($\approx10^{15}$\cmd) and $m$ is the mass of the
  adsorbing particle.
  Typically $\nu_0\sim 10^{12}$--$10^{13}$\s.
  The adsorption energies, $E_D$, of molecules onto the grains depends on the
  properties of the substrate.
  Table~\ref{table_ed} shows the adsorption energy, in K, of several
  fundamental species on different substrates that can be found in molecular
  clouds.
  Table~\ref{table_ed} also shows the evaporation time expected for each
  adsorption energy for two relatively close grain surface temperatures,
  $T_{\rm dust} =10$ and 15~K. It is easily visible the large differences
  between the different molecular species and substrates and the influence of
  temperature.

 \begin{table*}
   \centering
   \caption{Adsorption energies and evaporation timescales for different grain
     surfaces.}

   \begin{tabular}{lcccccccc}
     \hline\noalign{\smallskip}
     Species & H & \hd & \hd & He & O & CO & CO & CO \\
     Surface & H$_2$O ice & H$_2$O ice & \hd\ ice & Silicate & H$_2$O ice &
       graphite & H$_2$ O ice & CO ice \\
     $E_D$ (K) & 350$^a$ & 450$^{a,b}$ & 100$^c$ & 100$^a$ & 800$^a$ &
       1208$^d$ & 1740$^c$  & 960$^d$ \\
     \noalign{\medskip}
     \multicolumn{2}{l}{$T_{\rm dust}=10$ K} \\
     $t_{\rm evap}$ & 530 s & 0.46 yr & $\e{2}{-8}$ s & $\e{3}{-8}$ s &
       $\e{2}{15}$ yr & $\e{9}{32}$ yr & $\e{9}{55}$ yr & $\e{2}{22}$ yr \\
     \noalign{\smallskip}
     \multicolumn{2}{l}{$T_{\rm dust}=15$ K} \\
     $t_{\rm evap}$ & $\e{5}{-3}$ s & 4.5 s & $\e{7}{-10}$ s & $\e{1}{-9}$ s &
      $\e{4}{3}$ yr & $\e{3}{15}$ yr & $\e{6}{30}$ yr & $\e{2}{8}$ yr \\ 
     \noalign{\bigskip}
     \multicolumn{4}{l}{\textbf{On \hd\ coated grains}} \\
     \noalign{\smallskip}
     Species & CN & NH$_3$ & He & O & N$_2$ & CO & CO$_2$ \\
     Surface & \hd\ ice & \hd\ ice & Silicate & \hd\ ice & \hd\ ice & \hd\ ice
       & \hd\ ice \\ 
     $E_D$$^e$ (K) & 302 & 615 & 100 & 160 & 340 & 348 & 572 \\
     \noalign{\medskip}
     \multicolumn{2}{l}{$T_{\rm dust}=10$ K} \\
     $t_{\rm evap}$ & 24 s & $\e{1.7}{7}$ yr & $\e{3}{-8}$ s & $\e{2}{-5}$ s &
       18 min & 40 min & $\e{3.8}{5}$ yr \\  
     \noalign{\smallskip}\hline
   \end{tabular}

   \vspace{1.5mm}
   \begin{minipage}{16cm}
     \begin{list}{}{}
       \item[$^a$] from \citet{TielensAllamandola87}.
       \item[$^b$] 450 K was calculated by \citet{TielensAllamandola87} based
         on \citet{Hollenbach70}.
       \item[$^c$] from \citet{Sandford93}.
       \item[$^d$] from \cite{AllenRobinson77}. 
       \item[$^e$] Values estimated using a scaling factor of 1/5 of the
         adopted standard value of the adsorption energy, $E_D$
         \citep{AllenRobinson77}. 
         The original standard values of $E_D$ were: 
         CN on graphite, 1510 K \citep{AllenRobinson77}; \nht\ on \nht\ ice,
         3075 K \citep{Sandford93}; O on \hdo\ ice, 800 K
         \citep{TielensAllamandola87}; N$_2$ on \hdo\ ice, 1700 K
         \citep{TielensAllamandola87}; CO on \hdo\ ice, 1740~K
         \citep{Sandford93}; and CO$_2$ on \hdo\ ice, 2860 K
         \citep{Sandford93}.
     \end{list}
   \end{minipage}

 \label{table_ed}
 \end{table*}

  When modelling the accretion of gas-phase molecular species onto the grain
  surfaces, many adsorption energies are needed to describe the complex
  interacting system thus created \citep{Cuppen07}.
  Ideally, we would need to know the adsorption ener\-gies of the species onto
  the carbonaceous substrate, but also the energies between the species and
  ice (or different ice structures) and with other adsorbed species.
  Unfortunately, many of these ener\-gies are poorly known, if at all
  \citep[see][and re\-fe\-ren\-ces therein]{Cuppen07}.
  The lower half of Table~\ref{table_ed} also shows some estimates of the
  adsorption energies of several molecular species when the substrate is \hd.
  Most of these values were calculated assuming, for demonstration purposes and
  following \citet{AllenRobinson77}, a scaling value of 1/5 of the adopted
  standard value of $E_D$.
  The adsorption energies for \hd\ surfaces are considerably reduced when
  compared with the ones for silicate or \hdo\ ice surfaces.
  For ins\-tance, the adsorption energy for \hd\ on \hdo\ ice, $E_D=350$~K,
  while on \hd\ ice, $E_D=100$~K.
  The adsorption energies of other accreting heavy molecules behave in a
  similar way.
  Once $E_D$'s are reduced to 1/5 of the silicate surface values, evaporation
  timescales would be shorter than freezing timescales, $t_{\rm
    evap}\sim 40$ minutes for CO on \hd\ at 10~K (see Table 1).

\subsection{Coating of grains by \hd}
 \label{coating}

  By definition, \hd\ is the most abundant gas-phase species in mole\-cu\-lar
  clouds and several authors have long suspected that it could be an important
  component of the grain surfaces \citep{TielensHagen82,Govers80}.
  At the same time, the adsorption energies of \hd\ indicate that it is a
  rather volatile molecule.
  Table~\ref{table_ed} shows that the evaporation time for \hd\ on \hdo\ ice
  can go from a few seconds at a grain surface temperature of 15~K to 0.5
  years for 10~K.

  The surface coverage of \hd\ molecules will depend on the ba\-lance between
  their evaporation and accretion rates.
  The number of accreting \hd\ molecules per unit of time can be calculated
  from \citet{TielensHagen82}

  \begin{equation}
    R_{ac} = s_d \langle V_{\rm H_2}\rangle n({\rm H_2}) S = 0.10193 \left [
      \frac{n(H_2)}{10^4 {\rm cm^{-3}}} \right ] {\rm s^{-1}}
  \end{equation}
  where $n({\rm H_2})$ is the abundance of gas-phase \hd\ molecules, $\langle
  V_{\rm H_2}\rangle$ is the mean velocity of \hd\ molecules
  ($\sim1.026\times10^4 \sqrt{T_k}$ cm\s, where $T_k$ is the gas kinetic
  temperature), $s_d$ is the grain geometrical cross-section
  ($\sim3.14\times10^{-10}$\cmd), and $S$ is the sticking coefficient, which
  we will assume equal to 1.
  In practical terms, for a dense gas density, $\sim10^4$\cmt, a \hd\ molecule
  should hit a grain every 10 s.

  The number of \hd\ molecules that evaporate from the surface of a grain per
  unit of time can be expressed as

  \begin{equation}
    R_{evap} = N({\rm H_2})~ \nu_0~ {\rm exp} ( -E_D({\rm H_2}) / T )
  \end{equation}
  where $N$(\hd) is the number of \hd\ molecules on the grain surface.  

  If the number of grain sites, $N_s = 10^6$, the accretion time is $\sim10^7$
  s or $\sim0.3$ yr.
  For a dust temperature of 10~K, and $E_D({\rm H_2}-{\rm H_2O}) = 450$~K, the
  evaporation time is $\sim1.47\times10^7$ s or $\sim 0.5$ yr.
  The time needed for the system to reach steady state can be calculated as

   \begin{equation}
   \label{molh_abund}
    \frac{dN({\rm H_2})}{dt} = R_{ac}-R_{evap} = 0
  \end{equation}
  and the number of \hd\ molecules on the grain surface will be

  \begin{equation}
    N({\rm H_2}) = 1.5\times10^6 \left [ \frac{n({\rm H_2})}{10^4{\rm
          cm^{-2}}} \right ] 
  \end{equation}

  For a number of grain sites, $N_s$ = 1--3$\times10^6$, and $n$(\hd)
  $>10^4$\cmt, all the exposed sites will be occupied by \hd\ molecules in
  $\sim0.5$ yr at $n_{\rm H} > 10^4$\cmt\ \citep{TielensHagen82}.
  The evaporation of \hd\ from an \hd\ ice surface is very fast, leaving no
  possibility of multiple layers of \hd\ ice.

  Given the possibility that grains in molecular clouds are co\-vered by a
  layer of \hd\ molecules and, as Table~\ref{table_ed} shows, this can greatly
  reduce the adsorption energies of heavy molecules, it is re\-le\-vant to
  explore how much do the changes in the adsorption ener\-gies used in the
  chemical models change the values of molecular depletion in the gas.

\section{Results of the chemical modelling}

  We used our gas-grain chemical model based on the gas-grain chemistry of
  \citet{Hasegawa92} to explore the evolution of gas- and solid-phase
  abundances of species after assuming that the grains are coated with
  \hd\ molecules.
  The chemical model contained 3768 gas and surface reactions involving 503
  species, but we only took into account adsorption and evaporation of species
  from the grain surfaces, with no other chemical reaction on them apart from
  the formation of \hd.
  We ran several models at different gas densities ($2\times10^4$,
  $2\times10^5$, and $2\times10^6$\cmt, at a gas and dust temperature of 10~K,
  and cosmic ray ionisation rate $\zeta=1.3\times10^{-17}$\s.
  All models used initial elemental `low metal' atomic abundances
  \citep{Ruffle00}.

\begin{figure*}
   \centering
   \includegraphics[angle=-90,width=15cm]{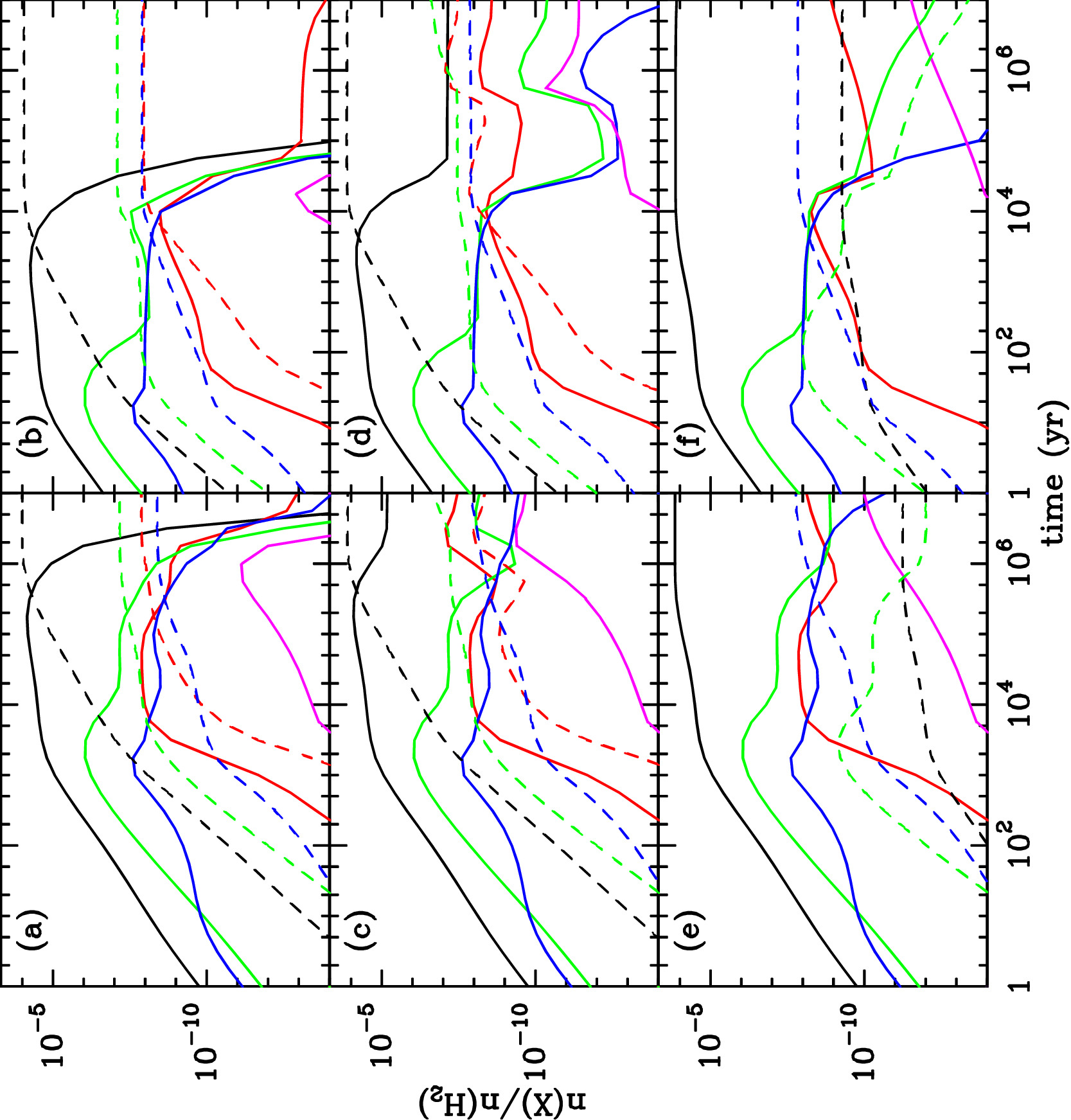}
   \caption{
     Abundances of gas-phase \textit{(solid lines)} and solid-phase
     \textit{(dashed lines)} CO \textit{(black)}, NH$_3$ \textit{(red)}, CN
     \textit{(green)}, CS \textit{(blue)}, N$_2$H$^+$ \textit{(purple)}, for
     the chemical models with $T_d$=10~K and:
     $(a)$ $n_{\rm H} = 2\times10^4$\cmt; $(b)$ $n_{\rm H} = 2\times10^6$\cmt;
     $(c)$ $n_{\rm H} = 2\times10^4$\cmt, 1/2$E_D$; $(d)$ $n_{\rm
       H}=2\times10^6$\cmt, 1/2$E_D$; $(e)$ $n_{\rm H} = 2\times10^4$\cmt,
     1/3$E_D$; $(f)$ $n_{\rm H} = 2\times10^6$\cmt, 1/3$E_D$.
   }
 \label{fig_models}  
 \end{figure*}

  We assumed a crude approach to simulate the effects of the coating of grains
  by \hd\ molecules and the subsequent modification of the adsorption
  energies, $E_D$.
  We ran three different sets of models:
  a fiducial one with no modification of $E_D$, and two mo\-dels where all
  $E_D$, except for H, \hd\, and He, were reduced by either a factor of 1/2 or
  1/3, a conservative value still far away from the scaling factor used by
  \citet{AllenRobinson77} to estimate the change in the adsorption energy of
  molecules landing on \hd-coated grains.
  Figure~\ref{fig_models} shows the results for two different number
  densities, $2\times 10^4$ and $2\times10^6$\cmt, and three different sets of
  adsorption energies: 
  standard values, 1/2 of the standard $E_D$ values, and 1/3 of the standard
  $E_D$ values.

  The upper row of Fig.~\ref{fig_models} shows the results of the fiducial
  case.
  For a gas density of $2\times10^4$\cmt, the gas-phase molecular abundances
  begin to suffer from depletion at a few times $10^5$ yr, at which point the
  gas and grain-phase abundances are approximately similar, and most of the
  molecular species disappear from the gas-phase at a few times $10^6$
  yr.
  This effect is accelerated by about a factor of 10 for a higher density,
  $2\times10^6$\cmt.

  The middle row of Fig.~\ref{fig_models} shows the results of the case where
  the adsorption energies are half of the standard values.
  There is still some degree of depletion (larger than 90\%) for some
  molecules, at a density of $2\times10^4$\cmt, but total depletion is never
  reached.
  Some molecules, such as \nht, do not have so high a depletion and the
  gas-phase abundance remains higher than the solid-phase abundance at all
  times.
  The level of depletion is much larger for higher densities, but even in this
  case there is not an almost total disappearance of the gas-phase molecules
  as in the standard case.
  On the other hand, the times when the solid-state abundances become larger
  than the gas-phase abundances do not seem to change that much.

  The case where the molecular adsorption energies have been reduced by a
  factor of 3 (bottom row of Fig.~\ref{fig_models}) shows the more dramatic
  changes.
  Many molecules, such as CO, CS (prime candidates for depletion) or \nht,
  show little or insignificant depletion, even at high densities, measured as
  the fraction of molecules in the gas and solid phases.
  This is not the case for CN, which still shows high depletion in this
  case. One must be cautious in this case, though, because the presence of CO
  in the gas at all times clearly affects the abundance of many molecules,
  such as CN and \ndh.

\section{Discussion and Conclusions}

  We have shown how the modelling of the depletion of heavy mole\-cular
  species in dense and cold regions of molecular clouds depends on the
  determination of accurate adsorption energies, $E_D$.
  Different values of $E_D$ affect the time molecular species reside on the
  grain surfaces and can provide a way to help explain the non-complete
  freeze-out of molecules from the gas phase.
  In particular, we argue that the adsorption energies of molecules on a
  substrate of \hd\ ice can be several times lower than the ones commonly used
  in current chemical models.
  We also showed that \hd\ could occupy all the available grain sites in a
  relatively short time, $\sim 0.5$ yr, which could produce a change in the
  adsorption rates of molecules on grain surfaces.

  We ran a few simple gas-grain models to test how much relatively small
  changes in the adsorption energies of molecular species affect the gas-phase
  abundances.
  We find that if we use values of $E_D$ half of the standard values, the
  depletion of most of the molecules is greatly reduced, although it is still
  at levels of a factor of 10.
  A further reduction of the molecular adsorption energies, down to a factor
  of $0.3 E_D$, which would be a conservative value of how much the adsorption
  energies can be changed, shows that molecules show very little or no
  depletion even at relatively high densities, $2\times10^6$\cmt.
  Thus, lower evaporation energies of heavy molecular species due to the
  coating of grains by \hd\ molecules may explain the presence of CO and other
  heavy molecules in the gas in average dark clouds.

  As we indicated in Sect.~\ref{coating}, we do not expect to have more than
  one monolayer of \hd\ molecules due to the almost instantaneous evaporation
  of any \hd\ landing on \hd-ice (see Table~\ref{table_ed}).
  On the other hand, we estimate that a CO molecule landing on a grain
  completely covered by \hd\ will be able to scan the $N_s$ sites on the
  surface of the grain in a time, $\sim 0.2$ s, much shorter than the expected
  evaporation time, $t_{evap}$(CO) $\sim40$ min.
  If the grain surface is only partially covered by \hd-ice, the effect will
  be that any CO molecule landing on \hd-ice will be able to find a free site
  outside of the \hd-ice, at which point it will remain trapped on the
  surface.
  This will reduce the number of available sites to be co\-ve\-red by
  \hd\ molecules by about 500 in a year, as we expect about one CO molecule
  landing on the surface per day at a gas density of $10^4$\cmt, and should
  not have any effect on the basic mechanism we discussed.

  In the calculations shown in Sect.~\ref{coating}, we assumed a ``flat'' or
  homogeneous grain surface, represented by just one value of the absorption
  energy, $E_D$, seen by the \hd\ molecules.
  But grain surfaces are thought to be rough, with an inhomogeneous structure
  of valleys, protrusions, cavities, and with and a certain degree of
  porosity.
  Landing \hd\ molecules will be then subject to a variety of binding energies. 
  Results of the effects of surface roughness on the formation rate of \hd\ on
  grains \citep{Cuppen05,Cuppen07,Cuppen06} show that irregularities affect
  the binding energies of the surface molecules, effectively increasing the
  binding energy ``seen'' by each molecule. 
  We expect that \hd\ molecules will tend to fill the valleys of the surface
  and, probably, help to build, locally, more than one layer of \hd\ ice.

  We realise that there are several objections that could be raised against
  this approach.
  There were some experimental results that show that the surface coverage of
  \hd\ might be $\sim 20$\% for a binding energy of 450~K
  \citep{Govers80,Schutte76}.
  Additio\-nally, in the simulations of \citet{Cuppen07}, \hd\ molecules pile
  up with CO in it.
  The potential dislocation of (almost completely) coating \hd\ by an incident
  \hd\ or heavier molecule can act against this proposed mechanism for
  desorption of heavy molecules from grains in dark
  clouds.
  \citet{TielensHagen82} also indicated that the \hd\ molecular layer may not
  be rigid enough to prevent oncoming heavier molecules to bind themselves to
  the surface underneath the \hd\ layer.

  Nonetheless, even if the \hd\ layer does not completely cover the grain
  surface and cannot completely avoid that landing heavy molecular species
  bind with the surface below it, there is no doubt that \hd\ molecules must
  be located all over the grain surfaces, due to the ubiquity of \hd\ in
  molecular clouds.
  This should have some, maybe substantial, effect on the effective adsorption
  energy value of molecular species that in the right environment makes this
  me\-chanism relevant.
  Particularly, after proving that some relatively small changes can affect a
  lot the depletion rate of heavy molecules.

  At the same time, the present mechanism and the several previously proposed
  desorption mechanisms could very well work simultaneously.
  It is yet to be seen which process dominates in a certain environment, how
  effective the \hd\ coating can be or in which kind of environments it would
  be dominant.

  The presence of a monolayer of \hd-ice on the surfaces of grains in
  molecular clouds could be observationally tested. 
  Unfortunately, there is no clear evidence yet of the detection of solid
  \hd\ on interstellar grains.
  \citet{SandfordAllamandola93} showed that solid \hd\ could be detected by an
  infrared absorption band at 2.417 $\mu$m (4137 cm$^{-1}$) attributed to the
  $Q_1$(1) pure vibrational transition of \hd.
  This band has never been detected and \citet{SandfordAllamandola93} argue
  that it would probably trace the \hd\ trapped in water ices.
  \citet{Schaefer07} proposed six pure $para$-hydrogen pair transition bands
  as candidates to explain unidentified emission features in ISO-SWS spectra
  of the NGC~7023 nebula.
  If any \hd-ice absorption band was detected, it would be a combination of
  existing surface and mantle solid \hd.
  Infrared spectroscopy studies of dust have found visual extinction
  thresholds of $\approx 3$ for \hdo\ ice \citep{Whittet88}, $\approx5$ for CO
  ice \citep{Whittet89}, and $\approx 4$ for CO$_2$ ice \citep{Whittet07},
  which roughly corres\-pond to very few monolayers of ice, probably between 2
  and 5 \citep{Hassel10}.
  The detection of only one monolayer of \hd\ ice, in the absence of any solid
  \hd\ in the mantle, may not be yet feasible.  

  Finally, it is interesting to consider if the \hd-coating mechanism proposed
  here can have any influence on the determination of the visual extinction
  thresholds discussed above.
  As Table~\ref{table_ed} shows, the evaporation time of \hd\ on \hd-ice is
  highly sensitive on tempe\-rature, $t_{evap}\sim 4.5$~s at 15~K.
  Thus, at the expected temperatures of the grains at those visual extinctions
  thresholds \citep[between 12 and 20~K,][]{DraineLee84,SmithSellgren93},
  \hd\ molecules do not have time to build up on the grain surfaces and we do
  not expect them to play any relevant role.
  Only when the grain temperatures go down to about 10~K will this mechanism
  delay the growth of ices

  Our simple modelling of the chemistry, changing the values of the adsorption
  energies, also proves how important is to obtain reliable estimates of
  evaporation energies, either laboratory measurements or theoretical
  calculations, for heavy molecules on \hd\ ice or, for that matter, in
  different kinds of substrate.
  A simplified experiment could be the study of \hd\ adsorption on various
  types of pure ice (\hdo, \nht, SO, SO$_2$, \hdco, ...) to estimate the
  interaction potential between the \hd\ molecules and other heavier
  molecules.

\section{acknowledgements}

  O.M.\ is supported by the NSC (Taiwan) ALMA-T grant to the Institute of
  Astronomy \& Astrophysics, Academia Sinica. 
  T.I.H.\ acknowledges the supports from NSC(Taiwan) 96-2112-M-001-018-MY3 and
  NSC 101-2911-I-001-503 (France-Taiwan, ORCHID)

\end{document}